\documentclass[pre,aps,amsmath,amssymb,eqsecnum,showkeys]{revtex4}
\usepackage{graphicx}
\newcommand{\xdif}{{\mathrm{d}}}
\newcommand{\Tr}{{\mathrm{Tr}}}
\newcommand{\iu}{{\mathtt{i}}}
\newcommand{\veps}{\varepsilon}
\newcommand{\bro}{\boldsymbol{\rho}}
\newcommand{\hm}{{\mathsf{H}}}
\newcommand{\pim}{{\mathsf{\Pi}}}
\newcommand{\um}{{\mathsf{U}}}
\newcommand{\calt}{\pmb{\mathcal{T}}}

\newcommand{\ca}{{\mathcal{C}}}
\newcommand{\tca}{\mathcal{R}}

\begin{document}
\clearpage
\preprint{}

\title{Quantum work fluctuations versus macrorealism in terms of non-extensive entropies}

\author{Alexey E. Rastegin}

\affiliation{Department of Theoretical Physics, Irkutsk State University,
Gagarin Bv. 20, Irkutsk 664003, Russia.}

\begin{abstract}
Fluctuations of the work performed on a driven quantum system can
be characterized by the so-called fluctuation theorems. The
Jarzynski relation and the Crooks theorem are famous examples of
exact equalities characterizing non-equilibrium dynamics. Such
statistical theorems are typically formulated in a similar manner
in both classical and quantum physics. Leggett--Garg inequalities
are inspired by the two assumptions referred to as the macroscopic
realism and the non-invasive measurability. Together, these
assumptions are known as the macrorealism in the broad sense.
Quantum mechanics is provably incompatible with restrictions of
the Leggett--Garg type. It turned out that Leggett--Garg
inequalities can be used to distinguish quantum and classical work
fluctuations. We develop this issue with the use of entropic
functions of the Tsallis type. Varying the entropic parameter, we
are often able to reach more robust detection of violations of the
corresponding Leggett--Garg inequalities. In reality, all
measurement devices suffer from losses. Within the entropic
formulation, detection inefficiencies can naturally be
incorporated into the consideration. This question also shows
advantages that are provided due to the use of generalized
entropies.
\end{abstract}


\maketitle

\pagenumbering{arabic}
\setcounter{page}{1}

\section{Introduction}\label{sec1}

In last years, the thermodynamics of quantum systems was a subject
of considerable research efforts in both theory and experiment.
Understanding of thermodynamic properties at the quantum level is
very important for several reasons. Quantum-mechanical laws should
be incorporated in a description of thermodynamics of
nanotechnology devices \cite{jareq11}. In emerging technologies of
quantum information processing, we have to control properly
interactions between quantum devices and their environment. A
fully quantum formulation of thermodynamic laws is still the
subject of active researches
\cite{horodecki13,ngour15,pnas15,csho15,rudolx15,amop16,modi16,ghrrs16}.
Some studies report that unitality replaces microreversibility as
the condition for the physicality of the reverse process in
fluctuation theorems \cite{albash,rast13,rkz14,aberg18}. Quantum fluctuation
theorems may allow us to understand
particle production during the expansion of the Universe
\cite{vedral16}. Possible applications of quantum theory to
information processing have renewed an interest to conceptual
questions. It turned out that some seemingly abstract concepts are
connected with practice more closely. One of quantum cryptography
protocols is directly connected with Bohm's reformulation
\cite{bohm51} of the Einstein--Podolsky--Rosen experiment
\cite{epr35}. Due to the celebrated work of Bell \cite{bell64},
quantum non-locality and contextuality still attract an
attention. Leggett--Garg inequalities \cite{lg85}
form one of directions inspired by Bell.

Various results concerning the question of Leggett and Garg are
reviewed in \cite{nori14}. In general, Leggett--Garg inequalities
\cite{lg85} are based on the two concepts known as the
macrorealism in the broad sense. First, one assumes that physical
properties of a macroscopic object preexist irrespectively to the
observation act. Second, measurements are non-invasive in the
following sense. In effect, performed measurement of an
observable at any instant of time does not affect its subsequent
evolution. These two assumptions are of classical nature and lead
to the existence of joint probability distribution for the
corresponding variables. In this regard, the consideration of
Leggett and Garg is somehow similar to the formulation inspired by
Bell. It is well known that restrictions of the Bell type can be
expressed in several different ways
\cite{shafiee05,bcpsw14,gisin14}. The very traditional form deals
with mean values of products of dichotomic observables
\cite{chsh69}. One of flexible and power methods is provided within
information-theoretic approach \cite{BC88}. Entropic formulation of
restrictions of the Leggett--Garg type was considered in \cite
{uksr12}. Further developments of such formulations in several
directions were accomplished in
\cite{rchtf12,krk12,rch13,rastqic14,wajs15,rastaop15,jia17}. In
particular, generalized entropies and related
information-theoretic functions have found use in these studies
\cite{rastqic14,wajs15,rastaop15}.

Experiments of the Leggett--Garg type concern the correlations of
a single system measured at different times. It turned out that
such results can well be applied to study fluctuations of the work
performed on a driven quantum system \cite{bm17}. This viewpoint
provides understanding of quantum fluctuation theorems from a new
perspective. The relations of Jarzynski \cite{jareq97a,jareq97b}
and Crooks \cite{crooks98,crooks99} are the first exact equalities
characterizing non-equilibrium dynamics. Such exact results were
originally formulated in the classical domain. Quantum versions of
fluctuation theorems are more sophisticated to express and
experimentally validate \cite{cht11}. Nevertheless, statistical
results of the discussed type are similarly formulated in
classical and quantum physics. To distinguish purely quantum
nature of work fluctuations in a driven quantum system, additional
concepts will be helpful. The simplest viewpoint on quantum work
is based on the projective measurement protocol. Note that a work
distribution exhibiting non-classical correlations may be measured
with a less-invasive coupling to a quantum detector \cite{sg2015}.
Such protocols lead to different work distributions and deviations
from the Jarzynski equality \cite{sg2015,sma2017}. The papers
\cite{hofer16,sg2016} proposed feasible experimental schemes to
detect negative quasiprobability of energy fluctuations in a
driven quantum system. It seems that the approach to quantum work
fluctuations on the base of Leggett--Garg inequalities has
received less attention than it deserves.

Dealing with work fluctuations, the authors of \cite{bm17}
utilized the dichotomic and entropic Leggett--Garg inequalities.
In the second case, they used formulation in terms of the Shannon
entropies. Applying statistical methods in numerous topics, some
extensions of the standard entropic functions were proposed. The
R\'{e}nyi \cite{renyi61} and Tsallis \cite{tsallis} entropies are
both especially important. The aim of this work is to apply
entropic functions of the Tsallis type to study quantum work
fluctuations within the macroscopic realism. The use of
formulation in terms of generalized entropies often allows us to
get more restrictive conditions and to widen a domain of their
violation \cite{rastqic14,rastaop15,rastctp14}. We show that this
conclusion also holds for testing quantum work fluctuations from
the macrorealistic viewpoint. The paper is organized as follows.
In section \ref{sec2}, we recall entropic formulation of
Leggett--Garg inequalities. Required properties of entropic
functions of the Tsallis type are given as well. In section
\ref{sec3}, we adopt $\alpha$-entropic Leggett--Garg inequalities
for quantum work fluctuations. Advances of the presented approach
to quantum work fluctuations versus the macrorealism are
exemplified in section \ref{sec4}. In particular, the use of
generalized entropies is physically important in a more realistic
situation with detection inefficiencies. In section \ref{sec5}, we
finish the paper with a summary of the results obtained.

\section{Entropic formulation of Leggett--Garg inequalities}\label{sec2}

In this section, we recall formulation of $\alpha$-entropic
inequalities of the Leggett--Garg type. Some results on entropic
functions of the Tsallis type will be listed. In general,
generalized entropic functions do not share all the properties of
the standard functions. In general, validity of desired properties
depend on the chosen values of the entropic parameter. These
points will be mentioned in appropriate places of the text.
Concerning the Bell and Leggett--Garg inequalities, we are mainly
interested in the chain rule and relations with respect to
conditioning on more.

Entropic formulations of non-contextuality and macrorealism are
interesting for several reasons. First, they can deal with
any finite number of outcomes. Second, entropic approach allows us
to concern more realistic cases with detection inefficiencies. Let
us consider a macrorealistic system, in which $X(t_{i})$ is a
dynamical variable at the moment $t_{i}$. Formally, the
macroscopic realism itself implies that outcomes $x_{i}$ of the
variables $X(t_{i})$ at different instants of time preexist
independently of their measurements. Further, the non-invasive
measurability claims that the act of measurement of the variable
at an earlier time $t_{i}$ does not affect its subsequent value at
a later time $t_{j}>t_{i}$. These assumptions inspire certain
corollaries \cite{kb08}. For each particular choice of time instants, the
statistics of outcomes is assumed to be represented by a joint
probability distribution $p(x_{1},x_{2},\ldots,x_{n})$. The joint
probabilities are then expressed as a convex combination of the
form \cite{kb12}
\begin{equation}
p(x_{1},x_{2},\ldots,x_{n})=
\sum\nolimits_{\lambda}
\varrho(\lambda)\,P(x_{1}|\lambda)\,P(x_{2}|\lambda)\cdots{P}(x_{n}|\lambda)
\, , \label{x1x2}
\end{equation}
where $\lambda$ denotes a collection of unknown ``hidden''
parameters. The right-hand side of (\ref{x1x2}) averages the
product of conditional probabilities $P(x_{i}|\lambda)$ by means
of hidden-variable probability distribution $\varrho(\lambda)$.
The latter remains to be unknown, but its existence {\it per se}
imposes certain restrictions. In any model, the probabilities
$P(x_{i}|\lambda)\geq0$ should obey
\begin{equation}
\sum\nolimits_{x_{i}}P(x_{i}|\lambda)=1
\, . \label{socn}
\end{equation}
Of course, hidden-variable probabilities $\varrho(\lambda)\geq0$
satisfy $\sum_{\lambda}{\varrho(\lambda)}=1$ as well. The above
formulas provide consistency conditions for a macrorealistic
model.

The existence of joint probability distribution (\ref{x1x2}) leads
to relations for conditional entropies. Entropic inequalities of
the paper \cite{uksr12} were obtained similarly to the treatment of
Braunstein and Caves \cite{BC88}. These inequalities were given in
terms of the corresponding Shannon entropies. There exist several
ways to extend standard entropic functions. Due to non-additivity,
the Tsallis entropies have found use in non-extensive
thermostatistics \cite{abe01,ar04,GMT04}. On the other hand,
entropic functions of the Tsallis type were fruitfully applied
beyond the context of thermostatistics. For instance, they allow
us to renew studies of some combinatorial problems \cite{ecount16}
and eigenfunctions of quantum graphs \cite{graphs17}. For basic
scenarios, inequalities in terms of non-extensive
entropies were derived in \cite{rastqic14}. In application to
macrorealistic models, this approach was accomplished in
\cite{rastctp14}.

Let us recall briefly the required definitions. Suppose that
discrete random variable $X$ takes its values according to the
probability distribution $\bigl\{p(x)\bigr\}$. For
$0<\alpha\neq1$, the Tsallis $\alpha$-entropy is defined by
\cite{tsallis}
\begin{equation}
H_{\alpha}(X):=\frac{1}{1-\alpha}\,\left({\sum\nolimits_{x}p(x)^{\alpha}}-1 \right)
 . \label{tsadf}
\end{equation}
With other denominator instead of $(1-\alpha)$, the entropy
(\ref{tsadf}) was proposed in \cite{havrda}. As a rule, the range
of summation will be clear from the context. If $Y$ is another
random variable, then the joint entropy $H_{\alpha}(X,Y)$ is
defined like (\ref{tsadf}) by substituting the joint probabilities
$p(x,y)$. Here, we follow chapter 11 of \cite{nielsen} in using
simplified notation for probabilities, so that
\begin{equation}
p(x)=\sum\nolimits_{y}p(x,y)
\, , \qquad
p(y)=\sum\nolimits_{x}p(x,y)
\, , \label{pxyj}
\end{equation}
and $p(x|y)=p(x,y)/p(y)$ due to Bayes' rule. It will be convenient
to use the $\alpha$-logarithm
\begin{equation}
\ln_{\alpha}(\xi):=
\begin{cases}
 \frac{\xi^{1-\alpha}-1}{1-\alpha}\>, & \text{for}\ 0<\alpha\neq1\, , \\
 \ln\xi\, , & \text{for}\ \alpha=1\,.
\end{cases}
\label{lanl}
\end{equation}
Then the right-hand side of (\ref{tsadf}) can be represented in a
more familiar form, namely
\begin{equation}
H_{\alpha}(X)=-\sum\nolimits_{x}p(x)^{\alpha}\ln_{\alpha}\bigl(p(x)\bigr)
\, . \label{tsadf1}
\end{equation}
The maximal value of $H_{\alpha}(X)$ is equal to
$\ln_{\alpha}(d)$, where $d$ is the number of different outcomes.
In the limit $\alpha\to1$, the Tsaliis $\alpha$-entropy is reduced
to the Shannon entropy
\begin{equation}
H_{1}(X)=-\sum\nolimits_{x}p(x)\,\ln{p}(x)
\, . \label{shadf}
\end{equation}
Fundamental properties of classical and quantum information
functions of the Tsallis type are discussed in
\cite{raggio,borland,abe02,abe04,sf04,dmb,sf06}. Some results
remain valid for more general families of entropies
\cite{hu06,rastjst11,bzhpl16}. Induced quantum correlation
measures were examined \cite{bbzpl16}. The R\'{e}nyi entropies
\cite{renyi61} form another especially important family of
parametrized entropies. Applications of such entropies in physics
are reviewed in \cite{ja04,bengtsson}.

In reality, detectors in measurement apparatuses are not ideal.
With respect to Bell inequalities, the role of this problem was
emphasized in \cite{shafiee05}. One of advantages of the entropic
formulation is that detector inefficiencies can easily be taken
into account \cite{rchtf12}. Let $\bigl\{p(x)\bigr\}$ denote the
original distribution of events in the experiment. We do not
actually deal with this distribution. To the given real
$\eta\in[0;1]$, we assign another probability distribution with
elements
\begin{equation}
p_{\eta}(x)=\eta\,p(x)
\, , \qquad
p_{\eta}(\varnothing)=1-\eta
\, . \label{peta}
\end{equation}
Here, the term $p_{\eta}(\varnothing)$ gives the probability of the
no-click event, whereas $\eta$ is the detector efficiency. The
probability distribution (\ref{peta}) corresponds to a
``distorted'' variable $X_{\eta}$ actually observed. To formulate
restrictions in terms of mean values, we have to put some
reference ``$\varnothing$-value'' of $X_{\eta}$. Otherwise, its mean
value is not defined. Such doubts are completely avoided within
the entropic formulation. For all $\alpha>0$, the entropy
$H_{\alpha}(X_{\eta})$ can be expressed as \cite{rastqic14}
\begin{equation}
H_{\alpha}(X_{\eta})=\eta^{\alpha}H_{\alpha}(X)+h_{\alpha}^{(b)}(\eta)
\, . \label{hxeta}
\end{equation}
Here, the binary $\alpha$-entropy is expressed as
\begin{equation}
h_{\alpha}^{(b)}(\eta):=
\frac{\eta^{\alpha}+(1-\eta)^{\alpha}-1}{1-\alpha}
\ . \label{bent}
\end{equation}
Let $Z=(X,Y)$ be an ordered pair of two random variables with the
joint probability distribution $\bigl\{p(x,y)\bigr\}$. In this
case, we consider a ``distorted'' variable $Z_{\eta\eta}$, for
which the probabilities are written as
\begin{align}
p_{\eta\eta}(x,y)&=\eta^{2}p(x,y)
\, ,&
p_{\eta\eta}(x,\varnothing)&=\eta(1-\eta)p(x)
\, , \nonumber\\
p_{\eta\eta}(\varnothing,\varnothing)&=(1-\eta)^{2}
\, ,&
p_{\eta\eta}(\varnothing,y)&=\eta(1-\eta)p(y)
\, , \label{peta2}
\end{align}
where $p(x)$ and $p(y)$ are defined by (\ref{pxyj}). For
$\alpha>0$, the entropy $H_{\alpha}(Z_{\eta\eta})$ reads as
\begin{equation}
H_{\alpha}(Z_{\eta\eta})=
\eta^{2\alpha}H_{\alpha}(Z)
+\eta^{\alpha}(1-\eta)^{\alpha}\bigl[H_{\alpha}(X)+H_{\alpha}(Y)\bigr]+h_{\alpha}^{(q)}(\eta)
\, . \label{hzeta}
\end{equation}
In the last formula, we used the quaternary $\alpha$-entropy expressed as
\begin{equation}
h_{\alpha}^{(q)}(\eta):=
\frac{\eta^{2\alpha}+2\eta^{\alpha}(1-\eta)^{\alpha}+(1-\eta)^{2\alpha}-1}{1-\alpha}
\ . \label{qent}
\end{equation}
Similarly to (\ref{hxeta}), the result (\ref{hzeta}) can be
checked immediately by substituting the probabilities
(\ref{peta2}) into the definition of the Tsallis $\alpha$-entropy.
We refrain from presenting the details here.

Entropic Leggett--Garg inequalities are conveniently formulated in
terms of the conditional entropy \cite{uksr12}. The entropy of $X$
conditional on knowing $Y$ is defined as \cite{CT91}
\begin{align}
H_{1}(X|Y):=\sum\nolimits_{y}p(y)\,H_{1}(X|y)
=-\sum\nolimits_{x}\sum\nolimits_{y}p(x,y)\,\ln{p}(x|y)
\, . \label{cshen}
\end{align}
One of the existing extensions of the conditional entropy
(\ref{cshen}) is posed as follows \cite{daroczy70}. Let us put the
particular function
\begin{equation}
H_{\alpha}(X|y):=-\sum\nolimits_{x}
p(x|y)^{\alpha}\,\ln_{\alpha}\bigl(p(x|y)\bigr)
\, , \label{pcen}
\end{equation}
which gives $H_{1}(X|y)$ for $\alpha=1$.
Then the conditional $\alpha$-entropy is defined as
\cite{sf06,rastkyb}
\begin{equation}
H_{\alpha}(X|Y):=\sum\nolimits_{y}
p(y)^{\alpha}\,H_{\alpha}(X|y)
\, . \label{qshen}
\end{equation}
It is easy to check that $H_{\alpha}(X|Y)\leq\ln_{\alpha}(d)$ for $\alpha\geq1$.

For the conditional $\alpha$-entropy (\ref{qshen}), the chain rule
takes place in its standard formulation. Due to theorem 2.4 of the
paper \cite{sf06}, for a finite number of random variables we have
\begin{equation}
H_{\alpha}(X_{1},X_{2},\ldots,X_{n})=
\sum_{j=1}^{n}H_{\alpha}(X_{j}|X_{j-1},\ldots,X_{1})
\, , \label{alchan}
\end{equation}
including
$H_{\alpha}(X,Y)=H_{\alpha}(Y|X)+H_{\alpha}(X)=H_{\alpha}(X|Y)+H_{\alpha}(Y)$.
For real $\alpha\geq1$ and integer $n\geq1$, the conditional
entropy (\ref{qshen}) obeys \cite{sf06,rastqic14}
\begin{equation}
H_{\alpha}(X|Y_{1},\ldots,Y_{n-1},Y_{n})
\leq{H}_{\alpha}(X|Y_{1},\ldots,Y_{n-1})
\ . \label{rlem1}
\end{equation}
Thus, conditioning on more can only reduce the $\alpha$-entropy of
degree $\alpha\geq1$. As exemplified in \cite{rastita}, this
property is generally invalid for $\alpha<1$. Studies of other
properties of generalized conditional entropies are reported in
\cite{sf06,rastita}.

Leggett--Garg inequalities in terms of Tsallis entropies are
posed as follows \cite{rastctp14}. Let $X_{j}$ be
shortening for $X(t_{j})$. We consider the case
$n=3$ with the variables $X_{1}$, $X_{2}$, $X_{3}$. For
$\alpha\geq1$, one gets
\begin{align}
H_{\alpha}(X_{1},X_{3})\leq{H}_{\alpha}(X_{1},X_{2},X_{3})
&=H_{\alpha}(X_{1})+H_{\alpha}(X_{2}|X_{1})+H_{\alpha}(X_{3}|X_{2},X_{1})
\nonumber\\
&\leq{H}_{\alpha}(X_{1})+H_{\alpha}(X_{2}|X_{1})+H_{\alpha}(X_{3}|X_{2})
\, . \label{cmb01}
\end{align}
A utility of this formulation in comparison with the usual one was
exemplified in \cite{rastctp14}. Note that the condition
$\alpha\geq1$ is caused by the property (\ref{rlem1}). We will use the above
formulation to study quantum work fluctuations from the viewpoint
of the macrorealism.

\section{Quantum work and characteristics of its distribution}\label{sec3}

In this section, we will discuss the concept of work performed on
a quantum system during some control process. Suppose that the
time evolution of the principal system is governed by the von
Neumann equation
\begin{equation}
\iu\hbar\,\frac{\xdif\bro(t)}{\xdif{t}}=
\bigl[\hm(\lambda),\bro(t)\bigr]
\, . \label{vneq}
\end{equation}
The system Hamiltonian $\hm(\lambda)$ depends on time through
variations of the control parameter $\lambda(t)$. According to
the protocol, the system will obtain or emit certain portions of
energy. To quantify the work performed, we measure the energy of
the system before and after realizing the protocol. Although this
treatment is most known in quantum settings, it is equally valid
in the classical picture. However, the latter case assumes
that the measured value is completely deterministic in each moment
of time. In the quantum case, outcomes are not only random but
also alter the current state of the system. In this sense, the
equation holds only between adjacent points, at which measurements
are carried out.

By $t_{0}$, $t_{1}$, $t_{2}$ and so on, we will mean the moments
at which the energy measurements are performed.  Further, we will
write the spectral decomposition in the form
\begin{align}
\hm_{0}\equiv\hm(\lambda_{0})
&=\sum\nolimits_{k}\veps_{k}^{(0)}\pim_{k}^{(0)}
\, , \label{spd0}\\
\hm_{1}\equiv\hm(\lambda_{1})
&=\sum\nolimits_{\ell}\veps_{\ell}^{(1)}\pim_{\ell}^{(1)}
\, , \label{spd1}
\end{align}
and similarly for other Hamiltonians. We will assume that the
eigenvalues are non-degenerate and the projectors are all of rank
one. This assumption is physically natural, since existing
symmetries of the studied system are rather broken through
interaction with environment. If the protocol starts with the
initial state $\bro_{in}$, then the outcome $\veps_{k}^{(0)}$
appears with the probability
$\Tr\bigl(\pim_{k}^{(0)}\bro_{in}\bigr)$. Due to the reduction
rule, the post-measurement state is represented as
$\pim_{k}^{(0)}$. Hence, we easily calculate the $\alpha$-entropy
$H_{\alpha}(E_{0})$. The time evolution between the moments
$t_{0}$ and $t_{1}$ is described by the operator
\begin{equation}
\um(t_{1}|t_{0})=
\calt\left\{\exp\biggl(-\iu\hbar^{-1}
\int\nolimits_{t_{0}}^{t_{1}}\hm(\lambda)\,\xdif{t}\biggr)\right\}
 , \label{um10}
\end{equation}
where the super-operator $\calt$ implies the chronological
ordering. The conditional probability
$p\bigl(\veps_{\ell}^{(1)}\big|\veps_{k}^{(0)}\bigr)$ is given by
the quantum-mechanical expression
\begin{equation}
p\bigl(\veps_{\ell}^{(1)}\big|\veps_{k}^{(0)}\bigr)=
\Tr\bigl(\pim_{\ell}^{(1)}\,\um(t_{1}|t_{0})\,\pim_{k}^{(0)}\,\um(t_{1}|t_{0})^{\dagger}\bigr)
\, . \label{cp10}
\end{equation}
Here, the post-measurement state is now represented as
$\pim_{\ell}^{(1)}$. Combining (\ref{cp10}) with the probabilities
of each $\veps_{k}^{(0)}$, we determine all the quantities
required to calculate $H_{\alpha}(E_{1}|E_{0})$ and
$H_{\alpha}(E_{1})$. Similarly to (\ref{cp10}), the conditional
probability $p\bigl(\veps_{m}^{(2)}\big|\veps_{\ell}^{(1)}\bigr)$
is written as
\begin{equation}
p\bigl(\veps_{m}^{(2)}\big|\veps_{\ell}^{(1)}\bigr)=
\Tr\bigl(\pim_{m}^{(2)}\,\um(t_{2}|t_{1})\,\pim_{\ell}^{(1)}\,\um(t_{2}|t_{1})^{\dagger}\bigr)
\, . \label{cp21}
\end{equation}
Further, we obtain the $\alpha$-entropies
$H_{\alpha}(E_{2}|E_{1})$, $H_{\alpha}(E_{2})$ and so on. To study
quantum work fluctuations from the viewpoint of the macrorealism,
we will also consider the interval between $t_{0}$ and $t_{2}$ as
an entire one. According to laws of quantum mechanics, the
corresponding conditional probabilities are expressed as
\cite{bm17}
\begin{equation}
p\bigl(\veps_{m}^{(2)}\big|\veps_{k}^{(0)}\bigr)=
\sum\nolimits_{\ell\ell^{\prime}}\,
\Tr\bigl(\pim_{m}^{(2)}\,\um(t_{2}|t_{1})\,\pim_{\ell}^{(1)}\,\um(t_{1}|t_{0})\,\pim_{k}^{(0)}
\,\um(t_{1}|t_{0})^{\dagger}\,\pim_{\ell^{\prime}}^{(1)}\,\um(t_{2}|t_{1})^{\dagger}\bigr)
\, . \label{cp20}
\end{equation}
These probabilities are used to calculate
$H_{\alpha}(E_{2}|E_{0})$. In the case of more than three
measurement moments, probabilities are written in a similar
manner. Calculating required entropic functions, we will be able
to check quantum work fluctuations on conformity with the
restrictions imposed by the macrorealism.

We shall characterize work fluctuation as
follows. Each concrete value
$w_{\ell{k}}^{(10)}=\veps_{\ell}^{(1)}-\veps_{k}^{(0)}$ is one of
possible ways to perform some work on the system. Taking into
account all the realizations, we then obtain the $\alpha$-entropy
\begin{equation}
H_{\alpha}(W_{10})=H_{\alpha}(E_{0},E_{1})=H_{\alpha}(E_{1}|E_{0})+H_{\alpha}(E_{0})
\, . \label{hw20}
\end{equation}
At the last step, the chain rule (\ref{alchan}) was applied. So,
the probability distribution for the work done on the system is
governed by the joint probability of two pertaining projective
measurements \cite{bm17}. This simplest viewpoint on work at the
quantum level is usually referred to as the projective measurement
protocol. Furthermore, the entropies $H_{\alpha}(W_{21})$ and
$H_{\alpha}(W_{20})$ are expressed similarly to (\ref{hw20}) by
appropriate substitutions. Suppose that the correlations
between physical quantities of interest are consistent with the
macroscopic realism. For $\alpha\geq1$, we rewrite the entropic
Leggett--Garg inequality (\ref{cmb01}) in the form
\begin{equation}
H_{\alpha}(W_{20})
\leq{H}_{\alpha}(W_{21})+H_{\alpha}(W_{10})-H_{\alpha}(E_{1})
\, . \label{algw}
\end{equation}
It turned out that quantum
work fluctuations can violate the above restriction. To
characterize the amount of violation, we introduce  the quantity
\begin{equation}
\ca_{\alpha}:=H_{\alpha}(W_{20})+H_{\alpha}(E_{1})-H_{\alpha}(W_{21})-H_{\alpha}(W_{10})
\, . \label{charal}
\end{equation}
The right-hand side of (\ref{charal}) can be rewritten in terms of
three conditional entropies, each of which does not exceed
$\ln_{\alpha}(d)$. The latter gives a natural entropic scale.
Hence, we will mainly refer to the rescaled characteristic
quantity
\begin{equation}
\tca_{\alpha}:=\frac{\ca_{\alpha}}{\ln_{\alpha}(d)}
=\frac{H_{\alpha}(W_{20})+H_{\alpha}(E_{1})-H_{\alpha}(W_{21})-H_{\alpha}(W_{10})}{\ln_{\alpha}(d)}
\ . \label{rcharal}
\end{equation}
For $\alpha=1$, this rescaling merely replaces the natural
logarithms with the logarithms taken to the base $d$.

The macrorealism in the broad sense demands that
$\ca_{\alpha}\leq0$ and $\tca_{\alpha}\leq0$ for all
$\alpha\geq1$. When one has observed strictly positive values, 
we conclude that quantum work fluctuations are not
consistent with the macrorealism. Setting $\alpha=1$, the result
(\ref{algw}) reduces to the inequality given in \cite{bm17}. In
this sense, we obtained a generalization of the previous result in
terms of non-extensive entropies. In general, an entropic
formulation of the Bell theorem provides necessary but not
sufficient criteria for consistency with local hidden-variable
models \cite{rch13}. Adding an appropriate randomness in
experimental settings, such inequalities could be treated as
sufficient \cite{rch13}. Using the entire family of
$\alpha$-entropic inequalities provides another way, which does
not require additional cost for related tuning of the experimental
setup \cite{rastqic14}. In the sense of characterizing quantum
work fluctuations, advances of the $\alpha$-entropic approach will
be discussed in the next section.

Real measurement devices are inevitably exposed to noise. Hence,
we should somehow address the case of detection inefficiencies.
Within the entropic formulation, we can use the model resulting in
(\ref{hxeta}) and (\ref{hzeta}). It will be assumed that detectors
at different points of the protocol have constant efficiency in
the range of interest. As work fluctuations are characterized by a
pair of energies, the corresponding $\alpha$-entropies are
modified according to (\ref{hzeta}). Instead of the theoretical
value (\ref{charal}), we actually obtain an altered one, viz.
\begin{equation}
\ca_{\alpha}^{(\eta\eta)}:=H_{\alpha}\bigl(W_{20}^{(\eta\eta)}\bigr)
+H_{\alpha}\bigl(E_{1}^{(\eta)}\bigr)-H_{\alpha}\bigl(W_{21}^{(\eta\eta)}\bigr)-H_{\alpha}\bigl(W_{10}^{(\eta\eta)}\bigr)
\, . \label{waral}
\end{equation}
Doing some calculations, we finally obtain
\begin{align}
\ca_{\alpha}^{(\eta\eta)}&=\eta^{2\alpha}\ca_{\alpha}-\Delta_{\alpha}(\eta)
\, , \label{wcaral}\\
\Delta_{\alpha}(\eta)&=\eta^{\alpha}\bigl[\eta^{\alpha}+2(1-\eta)^{\alpha}-1\bigr]H_{\alpha}(E_{1})
+h_{\alpha}^{(q)}(\eta)-h_{\alpha}^{(b)}(\eta)
\, . \label{deta}
\end{align}
The maximal efficiency $\eta=1$ gives $\Delta_{\alpha}(1)=0$, so that the quantity
(\ref{wcaral}) is reduced to (\ref{charal}). In the case of
standard entropic functions, the formulas
(\ref{wcaral}) and (\ref{deta}) read as
\begin{align}
\ca_{1}^{(\eta\eta)}&=\eta^{2}\,\ca_{1}-\Delta_{1}(\eta)
\, , \label{wcaral1}\\
\Delta_{1}(\eta)&=\eta(1-\eta)H_{1}(E_{1})+h_{1}^{(b)}(\eta)
\, . \label{deta1}
\end{align}
where we used $h_{1}^{(q)}(\eta)=2h_{1}^{(b)}(\eta)$ and 
$h_{1}^{(b)}(\eta)=\!{}-\eta\ln\eta-(1-\eta)\ln(1-\eta)$. Thus, actual results lead
to a quantity decreased in comparison with its idealized value.
This decreasing takes into account not only the factor
$\eta^{2\alpha}$ but also the second term in the right-hand side
of (\ref{wcaral}). To provide a robust detection of the violation,
this second term should be reduced in comparison with the first
one.

\section{Violation of entropic Leggett--Garg inequalities for work fluctuations}\label{sec4}

In this section, we study entropic Leggett--Garg inequalities for
some concrete physical examples. In principle, we should specify
not only the unitary operators that govern time evolution. One
also needs in exact transformations between orthonormal bases of
the Hamiltonians $\hm_{0}$, $\hm_{1}$, $\hm_{2}$ and so on. As was
noted in \cite{bm17}, for each time interval all the required
operations can be combined into a single unitary matrix. For the
interval between $t_{0}$ and $t_{1}$, matrix elements of the
corresponding matrix are written as
$\bigl\langle\veps_{\ell}^{(1)}\big|\um(t_{1}|t_{0})\big|\veps_{k}^{(0)}\bigr\rangle$.
Hence, we could accept some natural forms of such matrices.

Let us begin with the case of a single qubit. The most general
form of unitary $2\times2$ matrix is described in theorem 4.1 of
\cite{nielsen}. Following \cite{bm17}, we consider a particular
choice, namely
\begin{equation}
\um_{10}=
\begin{pmatrix}
    \cos(\theta_{10}/2) & -\sin(\theta_{10}/2) \\
    \sin(\theta_{10}/2) & \cos(\theta_{10}/2)
\end{pmatrix}
 . \label{uth10}
\end{equation}
For other intervals, the final matrices will be written similarly.
The angles $\theta_{\ell{k}}$ refer to the corresponding
intervals. Hence, they parametrize the conditional probabilities
leading to the joint probabilities. In studies of work
fluctuations, the initial state $\bro_{in}$ of the protocol is
chosen as a thermal equilibrium state. The inverse temperature is
taken to be proportional to $1/\Delta{E}$, where $\Delta{E}$ is
the gap between the ground and excited levels. We also suppose
that $\theta_{21}=\theta_{10}=\theta$. It turned out that the
results are periodic with respect to $\theta$ with the primitive
period $\pi$.

\begin{figure}
\centering \includegraphics[height=7.4cm]{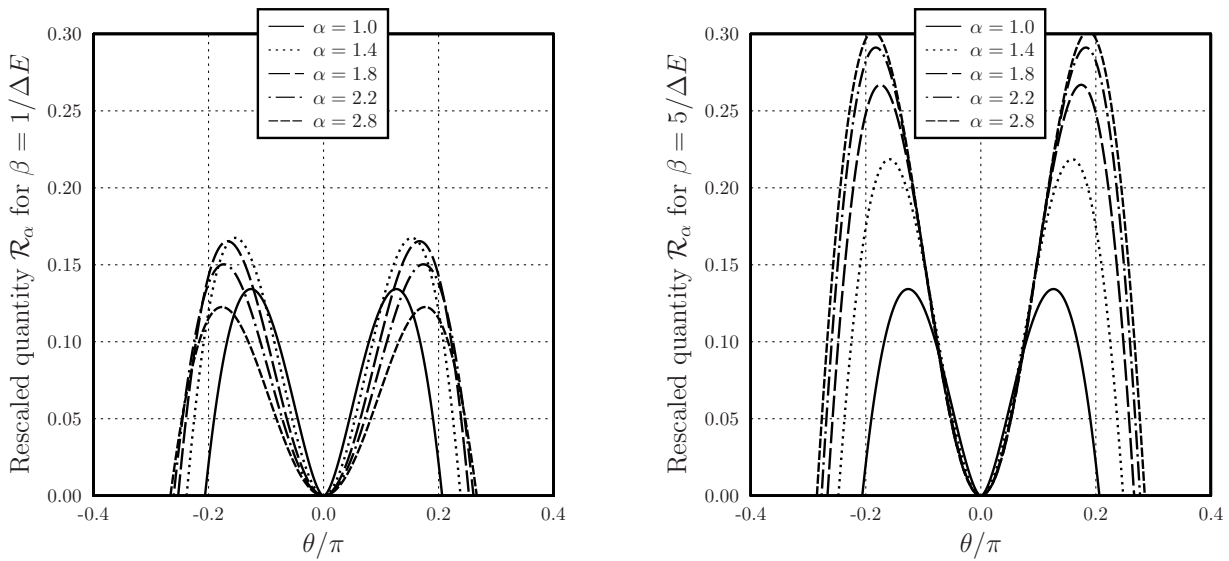}
\caption{\label{fig1} The rescaled characteristic quantity
(\ref{rcharal}) versus $\theta/\pi$ for the qubit case and several
values of $\alpha$, with $\beta=1/\Delta{E}$ on the left plot and
$\beta=5/\Delta{E}$ on the right one.}
\end{figure}

In the left plot of Fig. \ref{fig1}, we show the rescaled
characteristics (\ref{rcharal}) for $\beta=1/\Delta{E}$ and
several values of $\alpha$. The standard value $\alpha=1$
considered in \cite{bm17} is included for comparison. It is well
seen that the curve maximum goes to larger values of $\theta$ with
growth of $\alpha$. There exists some extension of the domain, in
which $\tca_{\alpha}>0$. For $\alpha\simeq2.6$, this extension
becomes negligible. Overall, the use of $\alpha$-entropies allows
us to widen the domain of visible violation by a range of order
$29.1$ \%. Both the strength and the range of violations can
actually be increased in this way. We also recall that the
considered situation corresponds to the choice
$\theta_{21}=\theta_{10}$. For other situations, the
$\alpha$-entropic approach may give additional possibilities to
test deviation of quantum work fluctuations from the macrorealism.

The authors of \cite{bm17} mentioned that, for a two-level system,
the violation of the entropic Leggett--Garg inequalities does not
depend on the initial state temperature. We have observed that
this is not the case with the use of $\alpha$-entropies. In the
considered model, this fact reflects the pseudo-additivity of the
Tsallis entropy. The pseudo-additivity is often quoted with
respect to non-extensive  composite systems, though such immediate
treatment was criticized \cite{aber01}. To illustrate this
finding, we present the rescaled characteristics (\ref{rcharal})
for $\beta=5/\Delta{E}$ in the right plot of Fig. \ref{fig1}.
Actually, for $\alpha=1$ the curve of (\ref{rcharal}) remains the
same as in the left plot. It is seen that both the strength and
the range of violations can be increased due to variations of
$\alpha\neq1$. Of course, the maximums of the curves depend on the
denominator used in (\ref{rcharal}). Even if we do not take into
account this point, the range of violations is really increased.
For the initial inverse temperature $\beta=5/\Delta{E}$, we add
the domain of visible violation by a range of order $34.3$ \%. The
latter is larger than the value of order $29.1$ \% that we
observed on the left.

In any case, the $\alpha$-entropic approach to work fluctuations
is sensitive to the temperature of a single qubit. That is, our
approach concerns one of genuine thermodynamic characteristics. We
should remember here that inequalities of the form
$\tca_{\alpha}\leq0$ do not guarantee that some probabilistic
model is consistent with the macrorealism in the broad sense. But
their violations clearly show that quantum work fluctuations are
not consistent with the macrorealism. It is natural to ask what
happens with $\tca_{\alpha}$ when the parameter $\alpha$ becomes
more and more. On the one hand, strictly positive values of
$\tca_{\alpha}$ may be observed for sufficiently large values of $\alpha$. On
the other hand, both the strength and the range of violations are
decreased with growth of $\alpha$. Maximal values of
$\tca_{\alpha}$ become very small, especially when the initial two
probabilities are close to each other. Say, for $\alpha=50$ and
$\beta=1/\Delta{E}$ the maximum of $\tca_{\alpha}$ is
approximately $0.36\cdot10^{-7}$. Thus, a choice of very large values of the
parameter $\alpha$ does not lead to new findings.

\begin{figure}
\centering \includegraphics[height=7.4cm]{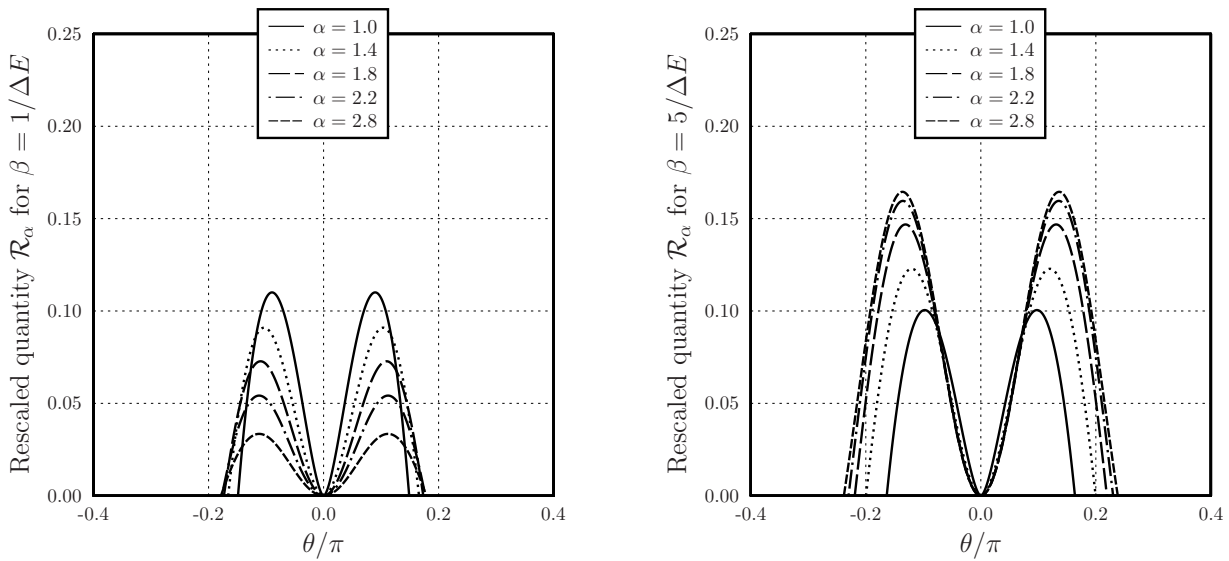}
\caption{\label{fig2} The rescaled characteristic quantity
(\ref{rcharal}) versus $\theta/\pi$ for the qutrit case and
several values of $\alpha$, with $\beta=1/\Delta{E}$ on the left
plot and $\beta=5/\Delta{E}$ on the right one.}
\end{figure}

Furthermore, we consider the case of a three-level system often referred to as a qutrit. One aims to show that a utility of the $\alpha$-entropic approach is not restricted to the qubit case. The qutrit model may be related to a system of two indistinguishable two-level particles, say, two spins in a magnetic field. Spin systems have found a considerable attention due to the so-called ``negative'' absolute temperature \cite{pound51,ramsey56}. Conceptual problems raised here are completely resolved within consistent thermostatistics \cite{dunkel14}. Assuming $\theta_{21}=\theta_{10}=\theta$, the corresponding unitary matrices will be taken as
\begin{equation}
\um_{21}=\um_{10}=\frac{1}{2}
\begin{pmatrix}
  1+\cos\theta & -\sqrt{2}\sin\theta & 1-\cos\theta \\
    \sqrt{2}\sin\theta & 2\cos\theta & -\sqrt{2}\sin\theta \\
  1-\cos\theta & \sqrt{2}\sin\theta & 1+\cos\theta
\end{pmatrix}
 . \label{uth210}
\end{equation}
It could be interpreted as the real rotation by $\theta$
about the axis given by the unit vector
$\bigl(1/\sqrt{2},0,1/\sqrt{2}\bigr)$. The three levels
are assumed to be equidistant with the gap $\Delta{E}/2$ between
two adjacent ones. The entropic quantities are then calculated
analogously to the qubit case. To facilitate the comparison, we
give Fig. \ref{fig2} in a similar manner. It presents the rescaled
characteristics (\ref{rcharal}) for $\beta=1/\Delta{E}$ in the
left plot and for $\beta=5/\Delta{E}$ in the right one. The
picture is like the qubit case in many respects. In particular,
the range of violations is increased by variations of
$\alpha\neq1$. Similarly to the qubit case, very large values of
$\alpha$ do not allow us to get new results including a width of
this range. So, we restrict a consideration to the values used in
Figs. \ref{fig1} and \ref{fig2}. In the right plot of Fig.
\ref{fig2}, the domain of visible violation is added by a range of
order $46.0$ \%. Again, a visible deviation
of work fluctuations from the macrorealism depends on the
temperature. Comparing Figs. \ref{fig1} and \ref{fig2}, we see
some reduction of the violation domain in the qutrit case.
Nevertheless, the presented results clearly show a utility of the
$\alpha$-entropic approach to detect deviation of quantum work
fluctuations from the macrorealism. Both Figures \ref{fig1} and
\ref{fig2} witness a connection between the inverse temperature
and the value of $\alpha$ corresponding to the maximal violation.
This interesting question is complicated to resolve.

\begin{figure}
\centering \includegraphics[height=7.4cm]{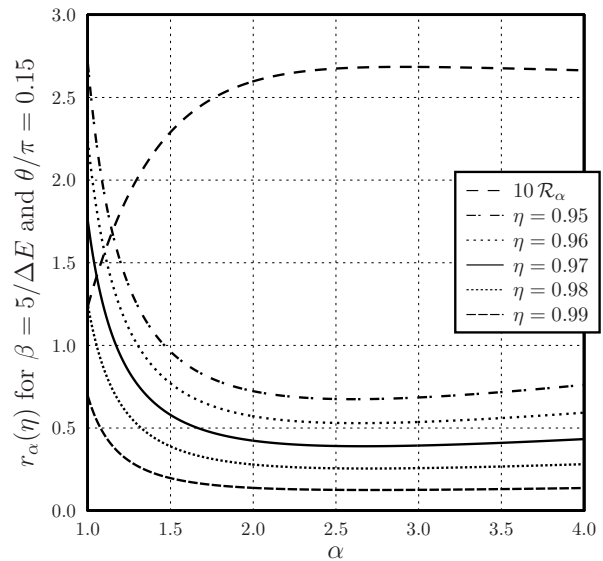}
\caption{\label{fig3} The ratio (\ref{ralt}) versus $\alpha$ for
the qubit case with $\beta=5/\Delta{E}$, $\theta/\pi=0.15$ and
several values of $\eta$. For convenience of referencing to Fig.
\ref{fig1}, one also shows values of $\tca_{\alpha}$ multiplied by
ten.}
\end{figure}

In reality, measurement devices are inevitably exposed to losses. In
particular, the detectors used are not perfect, so that the
no-click event will sometimes occur. Using the Shannon entropies,
the writers of \cite{rchtf12} considered the Bell inequalities in
the case of detection inefficiencies. For restrictions in terms of
generalized entropies, this question was examined in
\cite{rastqic14,rastctp14}. It turned out that the
$\alpha$-entropic approach allows us to reduce an amount of
required efficiency of detectors. The idealized characteristic
quantity $\ca_{\alpha}$ is given by (\ref{charal}). The macrorealism 
in the broad sense implies that $\ca_{\alpha}\leq0$
for $\alpha\geq1$. In the case of non-perfect detectors, we will
actually deal with the quantity (\ref{wcaral}). As was noted
above, we must confide that the violating term
$\eta^{2\alpha}\ca_{\alpha}$ is considerably large in comparison
with the reducing term (\ref{deta}). It is natural to inspect
their ratio
\begin{equation}
r_{\alpha}(\eta):=\frac{\Delta_{\alpha}(\eta)}{\eta^{2\alpha}\ca_{\alpha}}
\ , \label{ralt}
\end{equation}
which is restricted to the domain, where $\ca_{\alpha}>0$. Let us
consider this ratio in the qubit case with $\beta=5/\Delta{E}$.
Here, we set $\theta/\pi=0.15$, so that the strength of violations
is large for several values of $\alpha$ (see Fig. \ref{fig1} on the right). In
Fig. \ref{fig3}, we present $r_{\alpha}(\eta)$ versus
$\alpha\geq1$ for several values of $\eta$ close to $1$. For the
taken values of the parameters, a robust detection of violations
is hardly possible for $\eta<0.95$. If we restrict our
consideration to the standard entropies solely, then very high
efficiency is required. Using the $\alpha$-entropic approach
allows us to minimize (\ref{ralt}). As is seen in Fig. \ref{fig3},
this ratio essentially decreases with growth of $\alpha>1$. For
$\alpha\in(2;3)$, we are able to reach more reliable
detection of violations with the same measurement statistics. For
other choices of the parameters, a general picture is similar. In
any case, there are physical situations, in which variations of
the entropic parameter give additional possibilities to analyze
data about work fluctuations. In this sense, the $\alpha$-entropic
approach to quantum work fluctuations also deserves to be used in
studies of thermodynamics of small systems.

\section{Conclusions}\label{sec5}

We have studied quantum work fluctuations on the base of
$\alpha$-entropic Leggett--Garg inequalities. For all
$\alpha\geq1$, such inequalities express restrictions imposed by
the macrorealism on outcomes of measurements at different moments
of time. It turned out that microscopic ways to realize the
quantum work sometimes show statistical correlations inconsistent
with the macrorealism in the broad sense. Using the non-extensive
entropies allows us to reach additional possibilities in detecting
violations of Leggett--Garg inequalities. There are physically
motivated situations, in which the $\alpha$-entropic approach
leads to an extension of the domain of visible violations. Our
findings were illustrated with examples of a single qubit and a
single qutrit. As was mentioned in \cite{bm17}, for a qubit the
Leggett--Garg inequalities in terms of the Shannon entropies are
violated independently of the initial temperature. In opposite,
the $\alpha$-entropic approach to quantum work fluctuations versus
the macrorealism is sensitive to the temperature. Another reason
to use the $\alpha$-entropies concerns the case of detection
inefficiencies. This case can naturally be considered within the
entropic approach. Varying $\alpha>1$, we are sometimes able to
reduce essentially an amount of required efficiency of detectors.
Dealing with quantum work fluctuations, information-theoretic
functions of the Tsallis type should certainly be kept in mind.
Induced measures may give additional possibilities to examine
data, even though from time to time only.

\end{document}